\documentclass[reprint,prb,superscriptaddress,nobibnotes]{revtex4-1}
\usepackage{graphicx}
\usepackage[usenames,dvipsnames,svgnames,table]{xcolor}
\usepackage[amssymb]{SIunits}
\usepackage{amssymb}
\usepackage{latexsym,amsmath}

\begin{document}

\title{Transport through an impurity tunnel coupled to a Si/SiGe quantum dot} 

\author{Ryan H. Foote}
\email{rhfoote@wisc.edu}

\author{Daniel R. Ward}
\affiliation{Department of Physics, University of Wisconsin-Madison, Madison, Wisconsin 53706, USA}

\author{J. R. Prance}
\affiliation{Department of Physics, Lancaster University, Lancaster, UK}

\author{John King Gamble}
\author{Erik Nielsen}
\affiliation{Center for Computing Research, Sandia National Laboratories, Albuquerque, NM 87185, USA}

\author{Brandur Thorgrimsson}
\author{D. E. Savage}
\affiliation{Department of Physics, University of Wisconsin-Madison, Madison, Wisconsin 53706, USA}

\author{A. L. Saraiva}
\affiliation{Instituto de Fisica, Universitade Federal do Rio de Janeiro, Caixa, Postal 68528, 21941-972, Rio de Janeiro, Brazil}

\author{Mark Friesen}
\author{S. N. Coppersmith}
\author{M. A. Eriksson}
\email{maeriksson@wisc.edu}
\affiliation{Department of Physics, University of Wisconsin-Madison, Madison, Wisconsin 53706, USA}

\date{\today}

\begin{abstract}
Achieving controllable coupling of dopants in silicon is crucial for operating donor-based qubit devices, but it is difficult because of the small size of donor-bound electron wavefunctions.  Here we report the characterization of a quantum dot coupled to a localized electronic state, and we present evidence of controllable coupling between the quantum dot and the localized state.   A set of measurements of transport through this device enable the determination of the most likely location of the localized state, consistent with an electronically active impurity in the quantum well near the edge of the quantum dot.  The experiments we report are consistent with a gate-voltage controllable tunnel coupling, which is an important building block for hybrid donor and gate-defined quantum dot devices.
\end{abstract}

\maketitle

Donors in silicon are a natural choice for qubits \cite{Kane:1998p133}, because electron spins bound to donors have very long coherence times~\cite{Pla:2012p489,Tyryshkin:2012p143,Zwanenburg:2013p961}. Phosphorus donors also have nuclear spins with extremely long coherence times \cite{JohnJLMorton:2008p1458,McCamey:2010p1652,Muhonen:2014p986}. Although donor-based quantum devices can be fabricated with near-atomically precise placement of donors~\cite{Schofield:2003p136104,Fuechsle:2010p502}, even when well-placed, donors are very small, making it difficult to control and change the tunnel couplings between them with gate voltages.  In contrast, tunnel couplings are easily tunable in gate-defined quantum dots, and high-quality quantum dots hosting at least four different types of spin qubits have been demonstrated semiconductor materials~\cite{Petta:2005p2180,PioroLadriere:2008p776,Gaudreau:2011p54,Nowack:2011p1269,Maune:2012p344,Petersson:2012p380,Shulman:2012p202,Prance:2012p046808,Medford:2013p654,Kawakami:2014p666,Kim:2014p70,Shi:2014p3020,Veldhorst:2014p981}. An important feature of gate-defined quantum dots is that the electrons they contain can be displaced laterally simply by changing the voltages of the gates on the surface~\cite{Shi:2011p233108}.  Because of the ease of spatial control of the wavefunction in quantum dots, coupling quantum dots to localized states in semiconductors, if possible, would enable a path of a wide range of hybrid donor/quantum dot technologies.

In this letter, we report the observation of a controllable tunnel coupling between a localized electronic state and a gate-defined quantum dot formed in a Si/SiGe heterostructure.  We present measurements of transport through the device, demonstrating controllable tunnel coupling between the quantum dot and the localized state.  A set of stability diagram measurements enable a determination of the relative magnitude of the capacitance between the surface gates and both the quantum dot and the localized state.  We report the expected electron density profiles in the quantum dot and the neighboring reservoirs.  Combining the experimental results with 3D capacitive modeling based on the electron density profiles, we determine the most likely location of the localized state in the device.  The results reported here demonstrate that it is possible to control the tunnel rate between localized states and quantum dots, even though there is a dramatic difference in the characteristic length scale between these two electronic systems.

\begin{figure*}
\includegraphics[width=17cm]{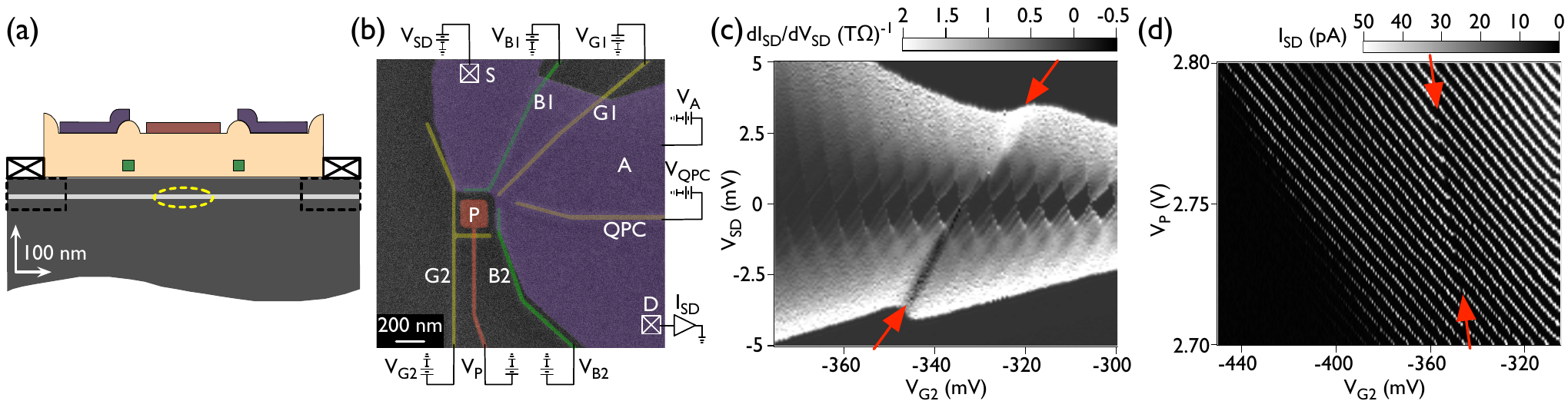}
\caption{\label{fig-1}Device design and characterization data of a quantum dot, revealing evidence of a nearby localized state. {\bf (a)} Schematic diagram showing a side view of the device. Substrate is a Si/Si$_{0.68}$Ge$_{0.32}$ heterostructure with a $\unit{10}\nano\meter$ silicon well (light grey) and $\unit{32}\nano\meter$ SiGe offset (dark grey). Both the upper (purple and red) and lower (green) layers of gates are $\unit{2}\nano\meter$ titanium and $\unit{20}\nano\meter$ gold deposited by electron beam evaporation. The lower gates were deposited on $\unit{10}\nano\meter$ of atomic layer depostion (ALD) grown aluminum oxide (light orange) while the upper gates were on $\unit{90}\nano\meter$ of oxide. Ohmic contacts S and D (denoted with $\boxtimes$ symbols) are $\unit{5}\nano\meter$ titanium and $\unit{40}\nano\meter$ gold on a region of the heterostructure degenerately doped with phosphorus through the quantum well (black dashed boxes).  Approximate location of the quantum dot and impurity are shown schematically in the center of the figure (yellow dashed oval). {\bf  (b)} Scanning electron micrograph of a completed device identical to the measured device.  Upper gate A (shown in purple) and paddle gate P (shown in red) were positively biased to accumulate a 2DEG in the reservoir and to control the energy of the dot, respectively.  On the lower level, gates G1, G2, and QPC (shown in yellow) were negatively biased to provide the dot confinement potential, whereas gates B1 and B2 (shown in green) controlled the tunnel barriers to the source (S) and drain (D) ohmic contacts (denoted with $\boxtimes$ symbols).  {\bf (c)} The derivative $dI_{\textrm{SD}}/dV_{\textrm{SD}}$ of the transport current with respect to the gate voltage $V_\textrm{G2}$.  Multiple Coulomb diamonds are observed. Near the center of the plot, a sharp resonance (indicated by arrows) is observed, suggesting a localized state. {\bf (d)} The current $I_\textrm{SD}$ at fixed $V_{SD} = \unit{100}\micro\volt$ as a function of $V_\textrm{P}$ and $V_\textrm{G2}$, showing many charge transitions of the dot. A jump in gate voltage of the location of the Coulomb blockade transitions is observed (indicated by arrows), corresponding to the localized state in (c).}
\end{figure*}

A gate-defined quantum dot was fabricated in a Si/Si$_{0.68}$Ge$_{0.32}$ heterostructure grown by chemical vapor deposition on a relaxed buffer layer in which the surface roughness had been removed by chemical-mechanical polishing.  A schematic diagram of a cross-section of the device is shown in Fig.\ \ref{fig-1}(a), and a top-view, false-color scanning electron micrograph of the sample is shown in Fig.\ \ref{fig-1}(b).  Measurements were performed in a dilution refrigerator with a mixing chamber temperature $T_{MC} < \unit{30}\milli\kelvin$.  

Figure~1(c) is a Coulomb diamond plot of the current through the device: it shows the derivative of the source-drain current between ohmic contacts S and D, as labelled in Fig.~1(b), as a function of the voltage $V_{G2}$ on gate G2.  The primary features in the plot are the diamond-shaped regions of very low differential conductance characteristic of Coulomb blockade with an average  charging energy $E_{c} = e^{2}/C _{Dot} =  \unit{760}\micro\electronvolt$.  By comparing this charging energy with the charging energies of few-electron Si/SiGe quantum dots~\cite{Simmons:2007p213103}, it is clear that this quantum dot is in the many electron regime.  We can also obtain, from the excited states visible in Fig.~1(c), an estimate of the single-particle energy of about $\unit{380}\micro\electronvolt$ in this quantum dot.  The data also enable the extraction of the proportionality constant (the lever arm) $\alpha_{G2} = \unit{148}\micro\electronvolt/\milli\volt$ between the voltage on gate G2 and the energy of the quantum dot.

In addition to these expected features, the red arrows near the center of Fig.\ \ref{fig-1}(c) highlight a sharp, isolated charging event.  Because this feature is so isolated---over a range in gate voltage corresponding to the addition of 13 electrons to the main dot only this one additional feature is observed---it corresponds to a very small, large charging energy object.  In addition, the capacitance between various gates and the object corresponding to this feature are different from those corresponding to the quantum dot.  This difference is made immediately clear in Fig.~\ref{fig-1}(d), which reports the current through the quantum device as a function of the voltages on gate P and  G2.  The phenomenology of this plot is very similar to those observed in metal-oxide-semiconductor devices in which donors have been implanted:~\cite{Morello:2010p687} near the center of the scan, a series of shifts in the charge transitions of the dot can be observed; these shifts correspond to the feature marked by the red arrows in Fig.~\ref{fig-1}(c).  The line through the gate voltage space spanned by $V_{P}$ and $V_{G2}$ connecting these shifts has a different slope than that of the Coulomb blockade peaks corresponding to the dot, confirming the presence of a nearby localized state that is not at the same physical position as the dot.

\begin{figure}
\includegraphics[width=8.5cm]{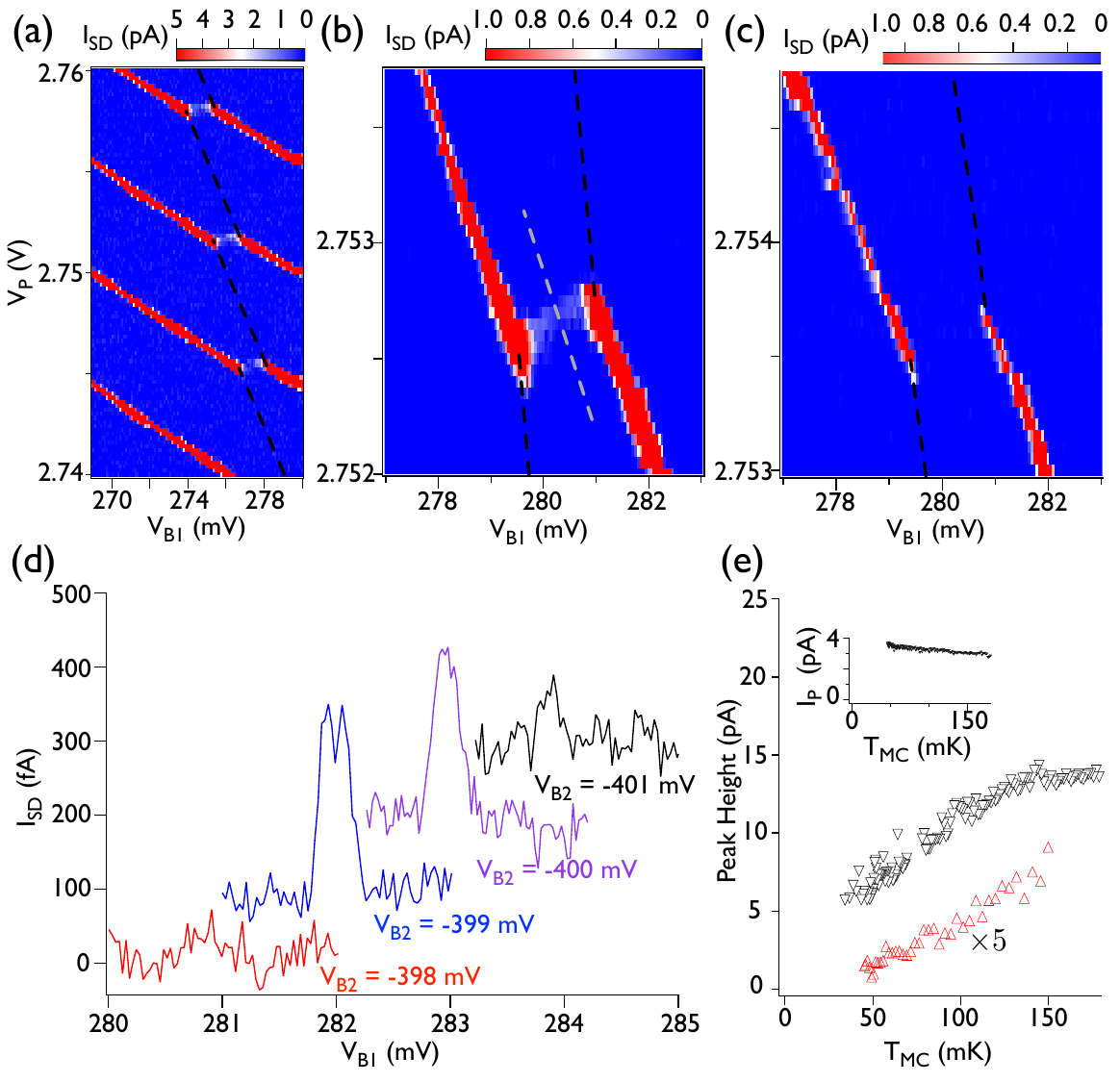}
\caption{\label{fig-2}Data demonstrating control of the tunnel coupling between the localized state and the quantum dot.  {\bf (a)} A high-resolution data set showing jumps in the Coulomb blockade transitions as a function of $V_\textrm{P}$ and $V_\textrm{B1}$.  A double dot-like stability diagram is revealed, including a slightly visible current on the polarization line.  The single electron transitions of the localized state are not visible in the current $I_\textrm{SD}$ and instead are shown schematically as black dashed lines. {\bf (b) \& (c)} Very high resolution scans of one of the state crossings shown in (a), for two different values of the gate voltage $V_{B2}$. In (b), $V_{B2}=-\unit{402}\milli\volt$ and current is present at the polarization line.  In (c), $V_{B2}=-\unit{404}\milli\volt$, and current is absent at the polarization line.  \textbf{(d)} Line scans along a path shown by the light gray line in panel (b).  The data for the $V_{B2} = -\unit{398}\milli\volt$ trace are not shifted, and subsequent traces are offset vertically by $\unit{100}\femto\ampere$ and laterally by $\unit{1}\milli\volt$ each for visibility.  These plots, which were acquired in close succession in time, show that changing $V_{B2}$ changes the tunnel couplings to the localized state, turning on and off current at the polarization line.  (The overall conditions in this plot are slightly different than those in panels b and c.)  {\bf (e)} Black inverted triangles show the height of the main Coulomb blockade peak near the anticrossing with the localized state, and red triangles show the height of the current peak on the polarization line, both of which are strongly temperature dependent.  \emph{Inset:} Coulomb blockade peak height $I_\textrm{P}$ far from the anticrossing with the localized state.  The temperature dependence in this regime is conventional for a large dot at reasonably low temperature.\cite{Houten:Book:1992}}
\end{figure}

Figure~\ref{fig-2}(a) reports a high-resolution measurement of the source-drain current $I_\textrm{SD}$ as a function of $V_{P}$ and $V_{B1}$.  The dashed black lines indicate the voltages at which the localized state charges, corresponding to the observed shift and gap in the Coulomb blockage peaks corresponding to the quantum dot.  No current is observed along these black dashed lines, indicating that---unlike the case for the quantum dot---the localized state is not tunnel coupled to both the source and the drain.  It is possible, however, that the localized state is connected to either the source or the drain, and this hypothesis is supported by the faintly visible line of current (dark blue in the color scale) that sits at the position of the polarization line in this stability diagram.

Figs.~\ref{fig-2}(b) \& (c) show  very high resolution measurements of a pair of triple points in this two-site system, for two different values of the voltage on gate B2.  In both plots, there is no current along the black dashed lines corresponding to the charge transition of the localized site, confirming that the localized site is not tunnel-coupled to both the source and the drain.  However, in Fig.~\ref{fig-2}(b), where $V_\textrm{B2} = -\unit{402}\milli\volt$, current is observed along the polarization line; in contrast, no such current is observed in Fig.~\ref{fig-2}(c), where $V_\textrm{B2} = -\unit{404}\milli\volt$.  This current is studied in more detail in Fig.~\ref{fig-2}(d), where we report line cuts across the polarization line, as indicated by the gray dashed line in Fig.~\ref{fig-2}(b).   As a function of gate voltage $V_{B2}$, Fig.~\ref{fig-2}(d) shows a dramatic evolution of the current along this path. For $V_{B2}$ equal to either -398 or -401~mV, no peak in current occurs at the polarization line.  In contrast, for intermediate values of $V_{B2} =$ -399 and -400~mV, a prominent peak in current is observed at the position of the polarization line.  The narrow voltage range of the barrier gate over which the tunneling condition is satisfied suggests that transport through the impurity is very sensitive to the relative strengths of the tunnel couplings of the dot and the impurity to each other as well as to the leads.

The difference between the current peak shown in Fig.~\ref{fig-2}(d) and the conventional Coulomb peaks corresponding to the quantum dot is also highlighted by the temperature dependence of each peak.  The inset to Fig.~\ref{fig-2}(e) shows the temperature dependence of the Coulomb blockade peak for the single dot, which is tunnel coupled to both the source and the drain, for gate voltages such that the localized state is not involved in the transport.  The current is nearly constant as a function of temperature, rising slightly as the temperature drops, consistent with a reasonably large quantum dot at temperatures $T$ for which $kT \ll E_c$ (Ref.~\onlinecite{Houten:Book:1992}).  In the main panel of Fig.~\ref{fig-2}(e), the black inverted triangles show the temperature dependence of the main Coulomb blockade peak very close to the anticrossing with the localized state, and the red triangles show the temperature dependence of the current peak on the polarization line.  In contrast with the behavior in the inset, both of these peaks increase strongly with increasing temperature, consistent with activated behavior.  This behavior is consistent with a localized state tunnel coupled to the dot and one (but not both) of the reservoirs.  Considering first the main Coulomb peak: on this peak, transport through the dot is allowed (by definition), but accessing the localized state requires thermal activation, which when present at elevated temperatures opens a parallel path to exit the dot, increasing the total current.  Considering the polarization line peak (which is more than 5 times weaker than the main Coulomb peak): this current is suppressed at low temperature.  Although charge can shuttle between the dot and the localized state at no energy cost, it cannot tunnel to or from the leads---raising the temperature activates this process.

\begin{center}
\begin{figure}
\includegraphics[width=8.3cm]{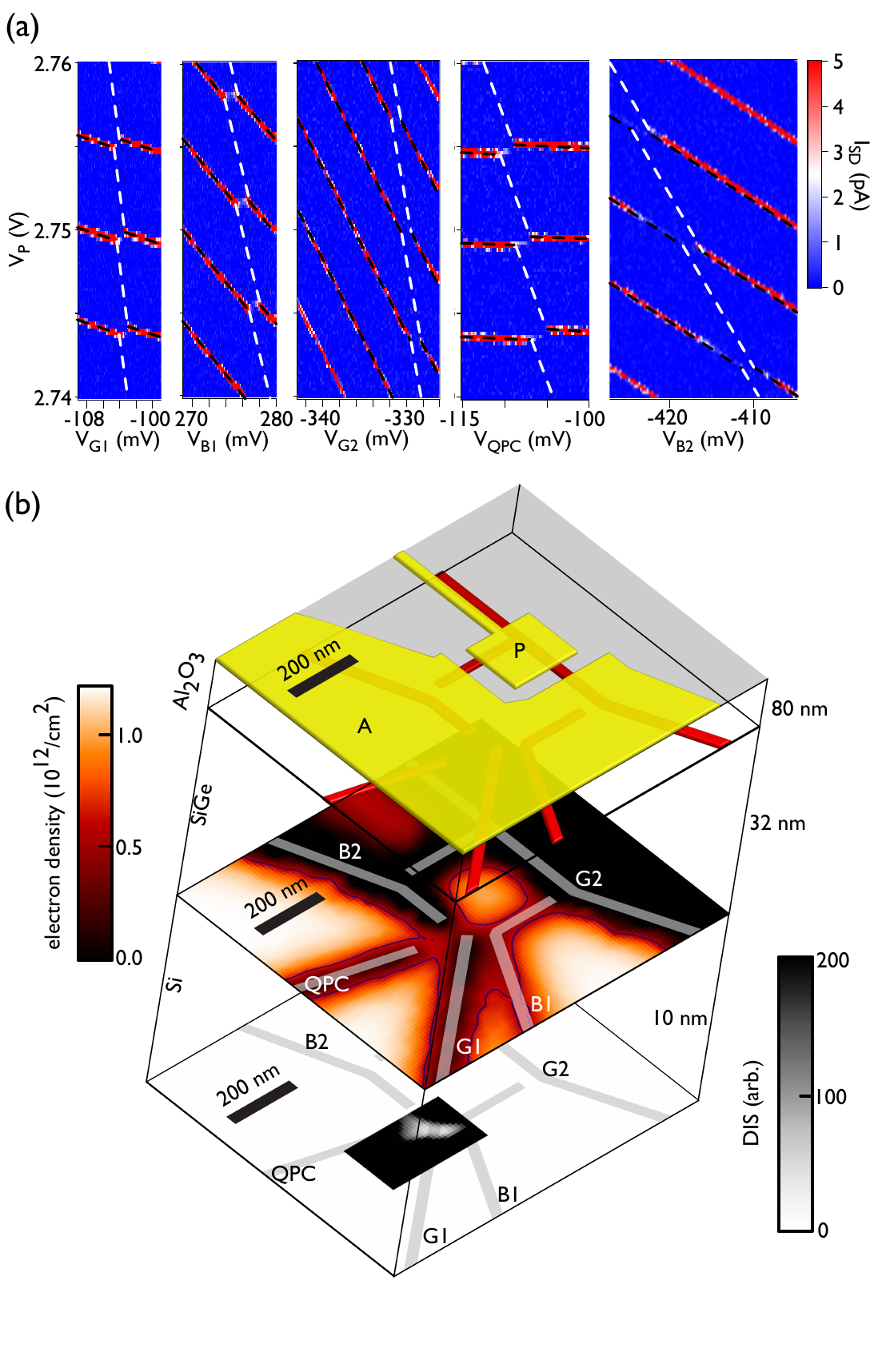}
\caption{\label{fig-3}Locating the localized state by combining experimental data and an electrostatic device model. {\bf (a)} A series of two-dimensional gate scans at fixed $V_{SD}$ measuring transport through the quantum dot-impurity stability diagram. Each plot reports the result of a different gate voltage being swept on the horizontal axis, versus the gate voltage $V_{P}$ on the vertical axis.  The black dashed lines highlight the Coulomb blockade transitions of the dot while the white dashed lines show the expected position of the unseen localized state charging event. \textbf{(b)} Results of 3D electrostatic modeling of the experiment to determine the location of the localized state.  The top layer shows the gate geometry of the device local to the dot. The middle layer shows the electron densities of the dot and leads as calculated by COMSOL, with the contour corresponding to $5\times10^{11}$ cm$^{-2}$ electron density.  
The bottom layer of the device shows the most likely location of the impurity, as determined by the discrepancy metric (Eq.~(\ref{eq-dis})), directly under the tip of gate G1 near the lower Si/SiGe interface about $\unit{10}\nano\meter$ below the dot and 2DEG.}
\end{figure}
\end{center}

To better understand these results, we combined systematic experimental stability diagram measurements with electrostatic device modeling to determine the location of the localized state~\cite{Mohiyaddin:2013p1903}.  Experimentally, we defined a central operating point in gate voltage space near the localized state-to-quantum dot charge transition.
Around this point, as shown in Fig.~\ref{fig-3}(a), we performed five two-dimensional stability diagram measurements, sweeping gate voltage $V_P$, which we use as our reference, as we stepped five other voltages: $V_{G1}$, $V_{B1}$, $V_{G2}$, $V_{QPC}$, and $V_{B2}$.  For each scan, all other voltages were held fixed to their values at the central operating point.

To interpret these data, we constructed an electrostatic device model in COMSOL Multiphysics\cite{Comsol}, using the device geometry from the experiment, as illustrated in Fig.~\ref{fig-3}(b).
The model was solved in the Thomas-Fermi Approximation\cite{Stopa:1996p13767,Schmidt:2014p044503}, with the self-consistent charge accumulation determined using a 2D density of states confined to a sheet at the Si-SiGe interface.
We assumed a $2\times2$ carrier degeneracy due to (2) spin and (2) valley degrees of freedom. 
The computational domain was $3\times 5$ $\mu$m laterally, and included a 100 nm air cap above the oxide layer and 500 nm SiGe substrate below the silicon well. 
We used zero-field boundary conditions on the sides of the domain and the top of the air cap, and set the conduction band edge to the Fermi level at the bottom of the domain.
In addition, we used voltages: $V_{G2} = -0.335$~V, $V_{B2} = -0.400$~V, $V_{QPC} = -0.100$~V, $V_{G1} = -0.104$~V,  $V_{B1} = +0.270$~V,  $V_{A} = +2.75$~V, and  $V_{P} = +2.75$~V.

We approximate the dot and 2DEG regions predicted by COMSOL as $\unit{5}\nano\meter$ thick metallic sheets at the $5\times10^{11}$ cm$^{-2}$
density contour (Fig.~\ref{fig-3}(b)), and we treat the localized state as a $\unit{1}\nano\meter$ radius metallic sphere.
For a given placement of the localized state, we construct a capacitance model that predicts each of the experimental charge-stability diagrams\cite{VanDerWiel:2003p1382}.
By rastering the localized state position across the device and computing as a fit metric a weighted sum-of-squared-differences between experimental and predicted values, we estimate the location of the localized state that is most consistent with the observed data shown in Fig. \ref{fig-3}(a). In particular, we sum the squared differences of two types of observed quantities:
\begin{enumerate}
\item{The slope of the line connecting all of the ``offsets" in the Coulomb blockade lines (the white dashed lines in Fig.~\ref{fig-3}(b)).\label{Sandia1}}
\item{The magnitude of the jump along the y-axis of a Coulomb blockade line due to the localized state.\label{Sandia2}}
\end{enumerate}
Quantities of type \ref{Sandia1} are unitless slopes whereas quantities of type \ref{Sandia2} have units of energy. To combine these into a single discrepancy metric, we found empirically that we needed to scale the type \ref{Sandia2} quantities reported in meV by $1 \times 10^{8}$ to balance them with the quantities of type \ref{Sandia1}. Thus, the overall discrepancy metric that is plotted in Fig.~\ref{fig-3}(b) is equal to:
\begin{equation}
DIS = \sum_{i=1}^{5} (PO_{i} -  EO_{i})^2 + 1\times 10^{8} \times \sum_{i=1}^{5} (PJ_{i} - EJ_{i})^2, \label{eq-dis}
\end{equation}
where $i$ indexes each of the five experimental slices shown in Fig.~\ref{fig-3}(a), $PO$ and $EO$ are the predicted and experimental type \ref{Sandia1} offset quantities, and $PJ$ and $EJ$ are the predicted and experimental type \ref{Sandia2} jump quantities respectively. The lower layer of Fig.~\ref{fig-3}(b) shows a cut of DIS along a plane 12~nm beneath the top of the strained Si well, identifying a region under the tip of gate G1 as the most likely region in the x-y plane to find the localized state. We find that the DIS value is not very sensitive to the depth (z-coordinate) between 10 and 20~nm, and thus we show the 12~nm data since this appears most consistent with the observed tunneling through the state.

Considering measurements and simulations together, we propose a tunnel rate dependent model of our hybrid quantum dot-impurity system.  Under typical device operation, the impurity is tunnel coupled to one of the leads and capacitively coupled to the dot.  Coupled this way, a charging event of the localized state varies the electric field local to the dot, changing the Coulomb blockade condition and resulting in the familiar jump in the dot charge transition from Fig.~\ref{fig-2}(c).  Changing the voltage of the tunnel barrier $V_{B2}$ changes the dot-drain, impurity-drain, and dot-impurity tunnel rates.  Under certain tunings, like those shown in Fig.~\ref{fig-2}(b), the three tunnel rates allow for current through the normally blockaded region as well as enhancement of current corresponding to the dot charge transitions.

In conclusion, we have shown measurements and modeling of a tunnel coupled quantum dot-impurity system in a Si/SiGe heterostructure.  We demonstrated tunable tunnel coupling between the impurity and the dot that is controlled by varying a nearby gate voltage, and we reported the temperature dependence of the coupled system.  We also have found the most likely position of the localized stated through capacitive modeling, and the capacitances extracted from this model are in good agreement with the experimental results.

The authors would like to thank Robert Mohr, Xian Wu, Dohun Kim, and Adam Frees for useful discussions.  Some features of the gate design used in this device are the topic of a patent application by M.A.E., J.K.G., D.R.W., S.N.C., and M.F.  This work has been supported in part by NSF (DMR-1206915, IIA-1132804), ARO (W911NF-12-1-0607) and the William F. Vilas Estate Trust.  Development and maintenance of the growth facilities used for fabricating samples supported by DOE (DE-FG02-03ER46028). This research utilized facilities supported by the NSF (DMR-0832760, DMR-1121288).  The work of J.K.G. and E.N. was supported in part by the Laboratory Directed Research and Development program at Sandia National Laboratories. Sandia National Laboratories is a multi-program laboratory managed and operated by Sandia Corporation, a wholly owned subsidiary of Lockheed Martin Corporation, for the U.S. Department of Energy’s National Nuclear Security Administration under contract DE-AC04-94AL85000.

\end{document}